# Artificial quantum confinement in LaAlO$_3$/SrTiO$_3$ heterostructures

M. Caputo[1], M. Boselli[2], A. Filippetti[3], S. Lemal[4], D. Li[2], A. Chikina[1], C. Cancellieri[5], T. Schmitt[1], J.-M. Triscone[2], P. Ghosez[4], S. Gariglio[2], V. N. Strocov[1]

[1]Swiss Light Source, Paul Scherrer Institute, CH-5232 Villigen, Switzerland

[2]Department of Quantum Matter Physics, University of Geneva, CH-1211 Geneva, Switzerland

[3]Department of Physics at University of Cagliari, and CNR-IOM, UOS Cagliari, Cittadella Universitaria, I-09042 Monserrato (CA), Italy

[4]Physique Théorique des Matériaux, Q-MAT, CESAM, Liège, B-4000 Liège, Belgium

[5]EMPA, Swiss Federal Laboratories for Materials Science & Technology, Ueberlandstrasse 129, Duebendorf CH-8600, Switzerland

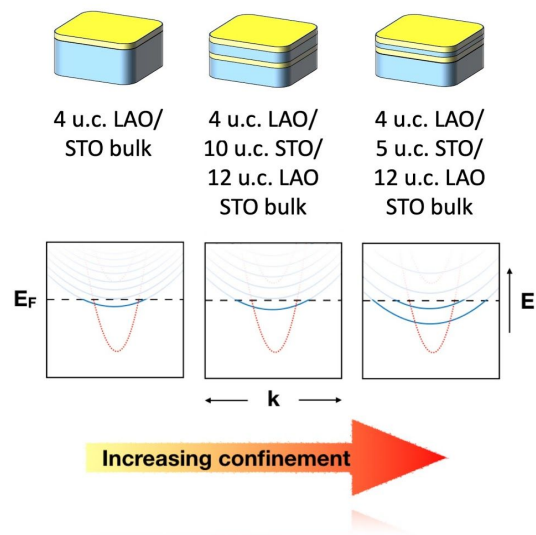

Heterostructures of transition metal oxides (TMO) perovskites represent an ideal platform to explore exotic phenomena involving the complex interplay between the spin, charge, orbital and lattice degrees of freedom available in these compounds. At the interface between such materials, this interplay can lead to phenomena that are present in none of the original constituents such as the formation of the interfacial 2D electron system (2DES) discovered at the LAO$_3$/STO$_3$ (LAO/STO) interface. In samples prepared by growing a LAO layer onto a STO substrate, the 2DES is confined in a band bending potential well, whose width is set by the interface charge density and the STO dielectric properties, and determines the electronic band structure. Growing LAO (2 nm) /STO (x nm)/LAO (2 nm) heterostructures on STO substrates allows us to control the extension of the confining potential of the top 2DES via the thickness of the STO layer. In such samples, we explore the dependence of the electronic structure on the width of the confining potential using soft X-ray ARPES combined with ab-initio calculations. The results indicate that varying the thickness of the STO film modifies the quantization of the 3d $t_{2g}$ bands and, interestingly, redistributes the charge between the $d_{xy}$ and $d_{xz}/d_{yz}$ bands.

Advances in thin-film deposition techniques have led to the discovery of a variety of phenomena in artificial heterostructures ranging from the quantum Hall effect to topological states. For epitaxial stacks, perovskites of transition metals are particularly interesting because, sharing a common crystal structure, they display a unique interplay of spin, orbital, charge and lattice degrees of freedom leading to a wide range of intriguing physical properties such as non-linear optical response, colossal magnetoresistance, high-temperature superconductivity, etc[1]. Among several phenomena reported for such heterostructures[2,3], one of the most remarkable observations is the conductivity discovered between two insulating materials. Originally detected at the interface between LaAlO$_3$ and SrTiO$_3$[3], two wide-gap insulators, it was also observed between LaTiO$_3$, a Mott insulator, and SrTiO$_3$[4] as well as between LaTiO$_3$ and KTaO$_3$[5]. The origin of the observed conductivity is related to the occurrence of a polar discontinuity between the two crystals, a common feature of these heterostructures[4,5]. Extensive work on the electronic properties of this interfacial system has shown that the band bending potential at the interface creates a quantum well (QW) that hosts the mobile electron system[6,7]. For interfaces prepared by growing a LaAlO$_3$ layer on top of a SrTiO$_3$ substrate, the vertical extension of the conducting region is found to be around 10 nm at low temperatures[8,9] and such confinement determines the electronic structure of the 2D electron system (2DES) - the $t_{2g}$ bands, degenerate in STO bulk, split into the $d_{xy}$ and $d_{xz}/d_{yz}$ ones as a consequence of the in-plane vs out-of-plane extension of these orbitals[10,7]. Although the polar discontinuity is absent at the bare STO surface, its certain treatments, for instance, UV illumination can dope electrons into the $t_{2g}$ manifold of the conduction band[11–13], inducing a metallic surface state which bears similarities to the LAO/STO interface.

The 2DES at the LAO/STO interface was found to be conducting, superconducting and to display a Rashba type spin-orbit interaction[5]. With this interface being a quite unique playground, several approaches have been used to manipulate its properties. The most common one is the field effect, particularly efficient due to the low density of the charge carriers, estimated on the order of a few $10^{13}$ cm$^{-2}$. Gating allowed, for instance, the tuning of the (super)conducting state[14,15], switching on and off the (zero-resistance) conductance, and the control of the spin-orbit interaction strength[16]. Other approaches have also been used to modify the properties of the 2DES such as variation of the growth conditions, stoichiometry, strain, oxygen deficiency, and absorption of gases on the LAO surface[17–19,20–22].

In this paper, we explore an interesting novel possibility to control the electronic properties of the 2DES. The idea is to induce the 2DES in an STO thin film whose thickness can be chosen below the natural extension of the 2DES in standard LAO/STO interfaces. Such an approach results in additional confinement of the 2DES which can modify its electronic structure. To this aim, we have grown a series of LAO/STO/LAO thin film heterostructures on STO substrates to control the extension of the QW potential through the thickness of the STO layer. Resonant ARPES and *ab-initio* calculations reveal the evolution of the electronic band structure of the 2DES with the STO layer thickness, uncovering a change in the energy separation and population of the $d_{xy}$ and $d_{xz}/d_{yz}$ bands as well as charge redistribution between them. These effects are driven by quantization of the electron states confined in the STO layer in the presence of electron correlations.

A sketch of the investigated structures is shown in Figure 1. On a (001)-oriented TiO$_2$-terminated STO substrate, a first LAO layer of 12 unit cells (u.c.) is grown by pulsed laser deposition. On top of this, we deposit the STO layer hosting the 2DES which is induced by a 4 u.c. LAO layer grown on top (details on the growth and physical properties of these samples, sensitive to the growth conditions, are provided in the Supporting Information).

Such structure creates 2DES localized in two regions, one at the top LAO film/STO film interface and one at the bottom LAO film /STO substrate interface. However, as explained below, ARPES is mostly probing the top 2DES whose vertical extension is limited by the thickness of the STO layer. The number ($n$) of unit cells for this STO layer was varied from $n=20$, a thickness corresponding to the self-confinement of the electron system that is observed in standard LAO/STO substrate interfaces at low temperature, down to $n=5$.

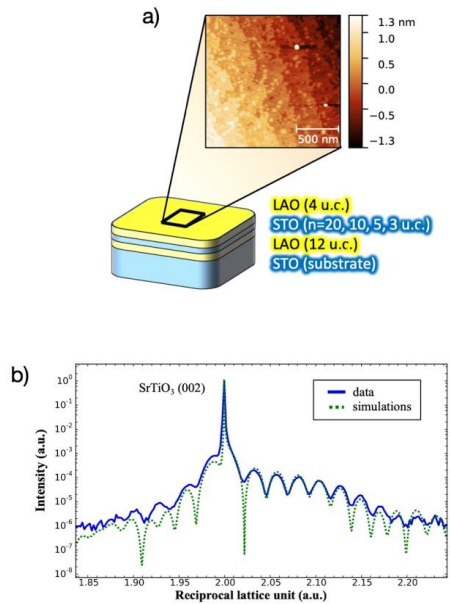

Figure 1: a) Scheme of the samples, and AFM image showing the typical topography of the topmost layer. b) XRD data of an example LAO (10 u.c.)/STO (35 u.c.)/LAO (5 u.c.) sample, showing the 002 reflection of the substrate together with the fringes originated from the ultra-thin films. Data are compared with a calculated pattern using the following out-of-plane axis values: c(LAO, 5u.c.)= 3.773 A, c(STO, 35u.c.)= 3.91 Å, c(LAO, 10 u.c.)= 3.766 A.

The electronic structure of these samples was investigated using (resonant) Soft X-ray Angle-Resolved Photoemission Spectroscopy (SX-ARPES) at the ADRESS beamline at SLS[23]. Using soft X-rays photons instead of the commonly used Vacuum Ultraviolet (VUV) ones, one can overcome the major drawbacks of the ARPES technique, namely, its extreme surface sensitivity. The probing depth of SX-ARPES is indeed higher by a factor of about four with respect to the VUV-ARPES[24], and its yield can be pushed even further using core-hole assisted photoemission processes[25]. Buried interfaces are a natural application field for this technique, and we refer the reader to a recent review for a detailed description[26]. The probing depth of SX-ARPES in our energy range accentuates the signal from the topmost LAO/STO interface, with the signal coming from the deeper second interface being less than 10% of the total for the ultimate limit of 5 u.c. of STO. All the data presented here, if not stated elsewhere, were acquired using a photon energy of 466 eV corresponding to the $L_2$ absorption edge of Ti: this gives us a clearer photoemission yield for the $d_{yz}$ band compared to the $d_{xy}$ one[25]. Moreover, the choice of s- polarized X-rays allows us to silence signal coming from the $d_{xz}$ band, due to the symmetry of our experimental geometry[23]. For the simulation of the electronic properties of the heterostructure, we use two beyond-standard methodologies which, correcting the known shortcomings of the standard Density Functional Theory, accurately describe strongly-correlated oxides: the variational pseudo self-interaction corrected (VPSIC) density functional theory,[27,28] and the B1-WC hybrid functional[29].

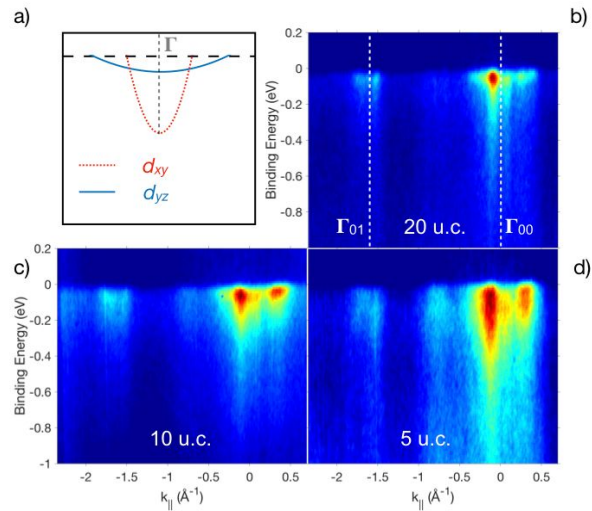

Figure 2: Panel a) shows a schematic of the electronic structure along the $\Gamma$-$X$ direction of the 2DES of a "standard" bulk LAO/STO interface. Panels b-d) are the experimental electronic structures for the $n$=20, 10 and 5 u.c. samples (marked on the panel).

Figure 2 shows experimental SX-ARPES intensity images representing the evolution of the electronic structure of the 2DES at the top LAO/STO interface of samples with a different thickness of the STO embedded layer. All the images are acquired in the same conditions along the $\Gamma$-$X$ direction of the Brillouin zone at a temperature of 13 K.

Figure 2b ($n$=20 u.c.) reveals an electronic structure very similar to the one of a "standard" LAO/STO interface realized at the $TiO_2$-terminated (001) STO substrate surface[30,31] (whose scheme is illustrated in Fig. 2a). At the Fermi level, the band structure is dominated by a weakly-dispersing band originated from the Ti $d_{yz}$ orbitals, extending through the region $k_{\parallel}= \pm 0.39$ Å$^{-1}$. Superimposed to this heavy band, two bright spots appear at $k_{\parallel}= \pm 0.09$ Å$^{-1}$: these correspond to the points where the band originating from the Ti $d_{yz/xz}$ orbitals hybridizes with the Ti $d_x$-originated orbitals. Strong out-of-plane localization of these bands makes its soft X-ray photoemission yield (and so its visibility) lower with respect to the deeper penetrating $d_{yz}$-derived bands[32]. Due to matrix element effects, the intensity is different between the two Brillouin zones ($\Gamma_{00}$ and $\Gamma_{01}$) and even not symmetric around the $\Gamma_{00}$ point ($k_{\parallel}= 0$ Å$^{-1}$), in particular on the right side of $\Gamma$ where the intensity of the $d_{yz}$ band is enhanced, while on the left side the intensity ratio between $d_{yz}$ and $d_{xy}$ band intensity is reversed.

Comparing these data with the measurements on the $n$=10 sample (panel c), we notice a clear enhancement of the photoemission yield of the $d_{yz}$ band, together with an energy broadening of the spectral structures. Interestingly, a sizeable spectral weight at the Fermi level appears in the region around -0.75 Å$^{-1}$, which falls on the boundary between the first and second Brillouin zone. We attribute this to an emerging 2x1 reconstruction already observed by Plumb *et al.*[33]: these authors suggest the occurrence of an octahedral tilting causing folding of the bands. Most importantly, the $n$=5 u.c. sample (panel d), shows a remarkably stronger broadening in the energy direction. In the following, we concentrate our discussion on this sample.

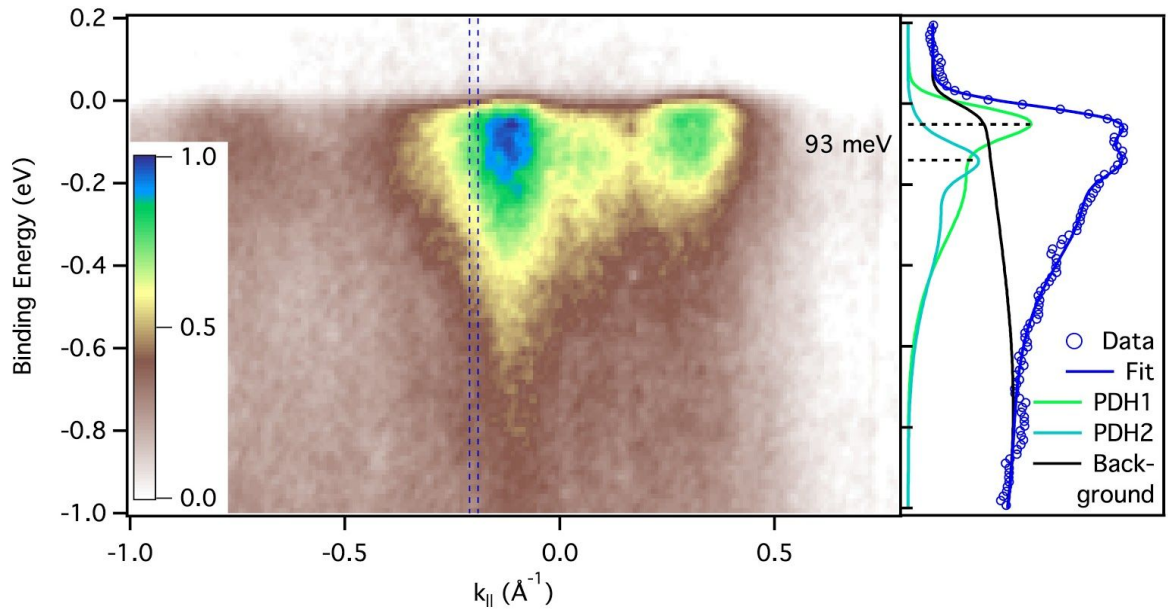

Figure 3: Detail of the electronic structure (left panel) of the *n*=5 u.c. sample along the **Γ-X** direction. The blue dashed lines represent the integration window of the EDC (right panel) plotted together with the fitting results. The horizontal dashed lines highlight the double-peak structure (see text for details).

In order to analyze in detail the spectral broadening for the *n*=5 u.c. sample, we plot in figure 3 an enlarged view of its electronic structure (left panel), together with an energy distribution curve (EDC) taken at $k_{||}$= -0.2 Å$^{-1}$ (right). The EDC (blue circles) clearly shows a double-peak structure in the range 0-0.2 eV of binding energy. The shape of the EDC is reproduced for all parallel momenta in the range -0.4 Å$^{-1}$ < $k_{||}$ < -0.2 Å$^{-1}$ and 0.2 Å$^{-1}$ < $k_{||}$ < 0.4 Å$^{-1}$ as shown in figure S1 (see Supporting Information), while for momenta in the range -0.2 Å$^{-1}$ < $k_{||}$ < 0.2 Å$^{-1}$ the presence of the $d_{xy}$ band introduces additional spectral weight that overshadows the double peak feature. The double-peak structure indicates the presence of a replica band, whose origin can be attributed to two phenomena: formation of a polaronic sideband due to enhanced electron-phonon coupling, or the quantization effect on the $d_{xy}$ band generating sub-band levels.

STO and STO-based interfaces are known to host polaronic charge carriers easily recognizable by a hump in the spectral function located roughly at 90-120 meV higher binding energy with respect to the quasiparticle (QP) peak[31,34,35]. This hump arises from electron coupling with the LO3 vibrational mode of the O cage. The double peak structure identified in our experimental EDC could in principle fit this picture, with the two peaks identifying the quasiparticle and its polaronic hump. In order to check this scenario, we fitted the EDC using a Gaussian peak for the QP band, followed by a series of Gaussians representing each of its polaronic replicas (see Supporting Information and figure S2 for details on the fitting procedure). Using just one of these peak-dip-hump (PDH) spectral shapes, it was not possible to fit the EDC's double peak structure. Moreover, we note that whereas all previous studies on the STO-based systems have found that the hump amplitude is significantly smaller compared to the QP peak[31,34,35], in our case the two spectral peaks have similar amplitudes. In contrast, a superposition of two PDH-structures yields a remarkable agreement with the experimental data, as shown in the right panel of Figure 3. The best fit is achieved for an energy separation between the two PDH curves of 95 ± 4 meV. The similar fitting procedure has been performed for several EDCs, resulting in a mean energy separation between the two bands of 99 ± 4 meV.

This analysis suggests that what we observe as the second spectral peak is actually a second $d_{yz}$ sub-band. We attribute the appearance of this band to a modification of the QW-potential hosting the 2DES: the confinement of the 2DES in the ultra-thin STO embedded layer of the LAO/STO heterostructure lowers the energy of the quantized second $d_{yz}$ sub-band below the Fermi level. In order to validate this scenario, in panel a) of Figure 4 we show the band structure calculated by VPSIC for a slab composed of 10 u.c. of STO with 4 u.c. of LAO layers on both STO sides, and a vacuum region equivalent to 3 u.c. LAO above. Since no 2DES has ever been observed when growing STO on LAO, presumably the bottom STO/LAO interface embeds a more shallow potential, making our calculation eventually a good approximation of the potential in two symmetric LAO/5 u.c STO interfaces. The calculations show two $d_{yz}$ bands separated by 116 meV, which is in good agreement with the experimental splitting reported in Figure 3. Note that the theoretical Fermi level position here is assigned by comparing calculations with the experimental bandwidth: in this case, the first QW-state resides completely below the Fermi level; this occurrence has already been observed by Wang *et al.* for the *n*=1 QW-state in anatase $TiO_2$[36], making this scenario plausible.

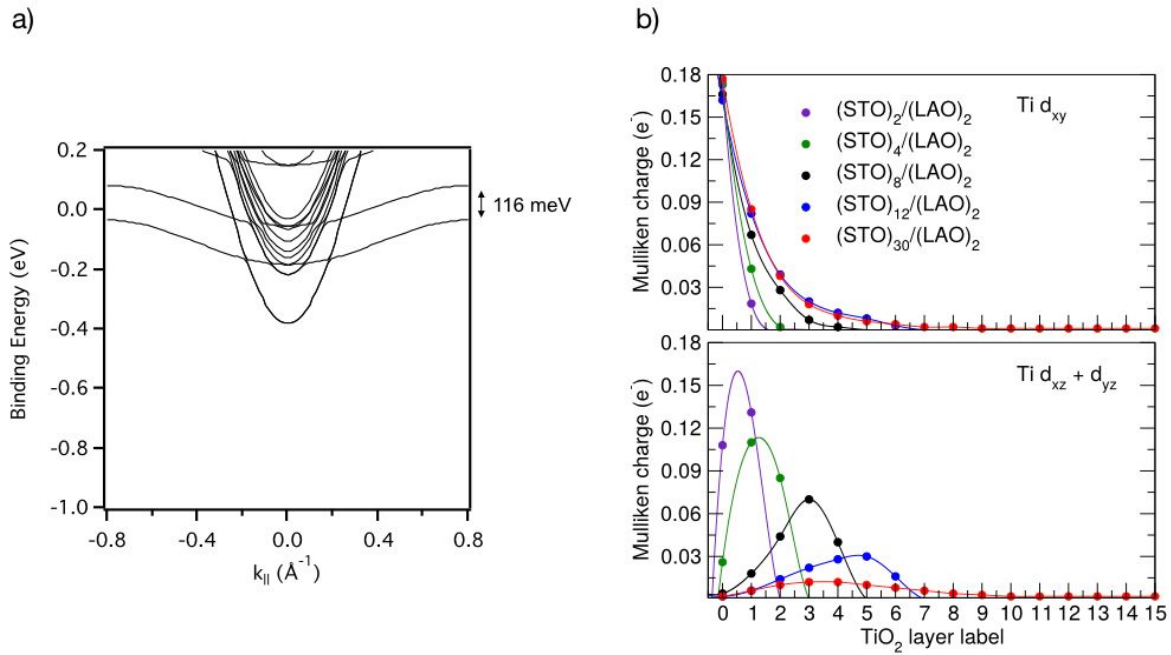

Figure 4: a) Band structure calculated for LAO/9.5 u.c. STO slab (see supporting info). b) Decomposition in the Mulliken scheme electron charge for different 2 u.c. LAO/ *n* u.c. STO superlattices (*n* = 2, 4, 8, 12, 30, see supporting info) in the different Ti sites. The labeling of the Ti sites starts at 0 for the first $TiO_2$ next to the LAO plane. Data are interpolated by the mean of cubic splines as a guide to the eye.

Calculations with the B1-WC energy functional considering a surface charge density of 0.5 e-/u.c. (i.e. the band filling sets the Fermi level self-consistently) confirm the shift of the $d_{yz}$ band to lower binding energy for ultra-thin STO layers. Panel b) in Figure 4 shows the Mulliken charges on each $TiO_2$ layer for STO slabs of different thickness, comparing the behavior of $d_{xy}$ orbital states (top graph) with one of $d_{xz/yz}$ states (bottom graph). When the

STO thickness is reduced, one can see, both in the top ($d_{xy}$ states) and in the bottom ($d_{xz/yz}$ states) plots, a progressive shift of the charge toward the interface, a natural effect of the thinning of the STO layer. More interesting, by comparing the two graphs, one can notice a charge transfer from the $d_{xy}$ to the $d_{xz/yz}$ states along with the decrease of the STO thickness: in the ultra-thin limit, the amount of charge accumulated in the two bands becomes nearly equal. This finding supports the idea that in the narrow STO slab limit a consistent fraction of charge can also populate the n=2 QW-state with the $d_{xz/yz}$ orbital character. This behavior can be attributed to two effects: on one hand, reducing the QW width increases the energy splitting between sub-bands of same symmetry, whereas, on the other hand, as the charge accumulates closer to the interface, the occupancy of the $d_{xz}/d_{yz}$ states, localized farther from the interface, is favored because their energies are lowered due to static correlation effects[37,38,39]. These effects, resembling Hubbard's *U*, are included into the exchange-correlation potential of our DFT calculations. These results indicate that a combination of quantum confinement with electron correlations, controlled by the thickness of the embedded STO layer, can shift the Fermi level of the 2DES and thereby give a control of the charge carriers density.

A consequence of the reduction of the QW width is the increase of the energy band spitting, appearing in the electronic structures presented in Figure S4 (see Supporting Information). Observation of such trend experimentally is however difficult due to the lack of energy resolution for the thicker STO layers, where the quantum wells are very close in energy, and because of the smearing of all spectral structures induced by the growing disorder and interfacial intermixing in heterostructures with ultra-thin STO layers. Figure S3 shows the experimental data for the *n*=3 sample: one can notice a qualitative agreement with the VPSIC calculations, but the experimental smearing hinders a more quantitative comparison.

In conclusion, the growth of artificial LAO/STO heterostructures with an embedded STO layer forms a 2DES with tunable properties. The tunability is achieved by varying the thickness of the STO layer on a single unit cell scale, which controls the confinement of the 2DES and thus energy position and population of different QW-states. In a wider perspective, the out-of-plane confinement at the interface can be complemented by confinement in the lateral direction, in order to obtain conducting channels with desired charge carriers density. This work paves the way for further nanoscale manipulation of 2DESs at surfaces/interfaces of different TMOs.

**Acknowledgment**

This work was supported by the Swiss National Science Foundation through Division II and by the European Research Council under the European Union's Seventh Framework Program (FP7/2007-2013)/ERC Grant Agreement no. 319286 (Q-MAC).

# Supporting information

**Reproducibility of the double peak feature** - The EDC shown in figure 3 of the main text is reproducible for all the parallel momenta in the range -0.4 Å$^{-1}$ < $k_{\parallel}$ < -0.2 Å$^{-1}$ and 0.2 Å$^{-1}$ < $k_{\parallel}$ < 0.4 Å$^{-1}$ as shown in figure S1, while for momenta in the range -0.2 Å$^{-1}$ < $k_{\parallel}$ < 0.2 Å$^{-1}$ the presence of the $d_{xy}$ band introduces additional spectral weight that overshadow the double peak feature.

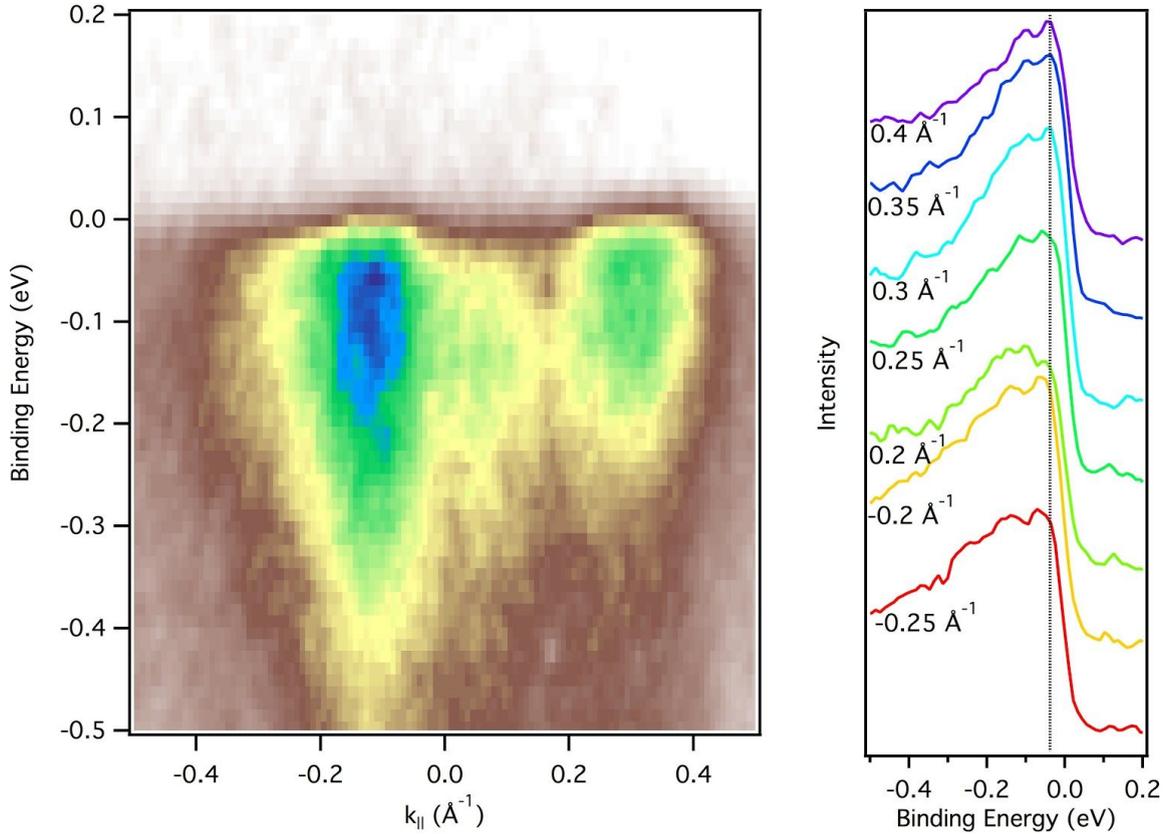

Figure S1: The left panel shows an enlarged view of the 2DES for the *n*=5 u.c. sample, while on the right panel the respective EDCs for different $k_{\parallel}$ values. For all of them the double peak structure underlined in the main text is evident. The vertical dotted line is a reference to make visible the dispersion of the bands.

**Fitting procedure** - Simplistically, the high-energy polaronic tail can be imagined by the superposition of a series of replica of the quasiparticle, each of them separated by $\omega_0$ (the frequency of the phonon mode coupled with electrons). This is analogous to the case of the independent oscillator model, where the intensity of the n-th replica follows the Poisson distribution $I_n = \frac{g^n}{n!} I_0$, being *g* a coupling constant. This description holds if a series of conditions can be applied to the system: a) Dispersionless phonon, b) Excitation energy such that $E-E_F \ll \omega_0$, c) Electronic bandwidth $W \ll \omega_0$, d) Temperature $T \ll \omega_0$. Previous application of this simplified model in the case of bare STO (ref 26, 27 in the main text) described successfully the observed lineshape.

We fitted EDC with two PDH structures: each of these is composed by a Lorentzian QP peak centered in a different $x_0$ and scale parameter $\gamma$, plus four additional Gaussian peaks each of them centered in $x_0+n\omega_0$ and variance $\sigma$. The coupling constant *g* is the same for the two polaronic humps, as well as the QP/hump intensity ratio. The fitting function is completed by a Gaussian peak accounting for the background and in-gap states situated in the region 0.5-1.5 eV of binding energy, and another Lorentzian peak accounting for the Ti $2p_{1/2}$ core level excited by the second order light coming from the beamline ($h\nu$=932 eV), and visible right after the Fermi level. Apart this last peak everything is multiplied by a Fermi Dirac function. The whole fitting function is convoluted with a Gaussian peak accounting for the finite experimental resolution. In figure S2 all the components of the fitting function.

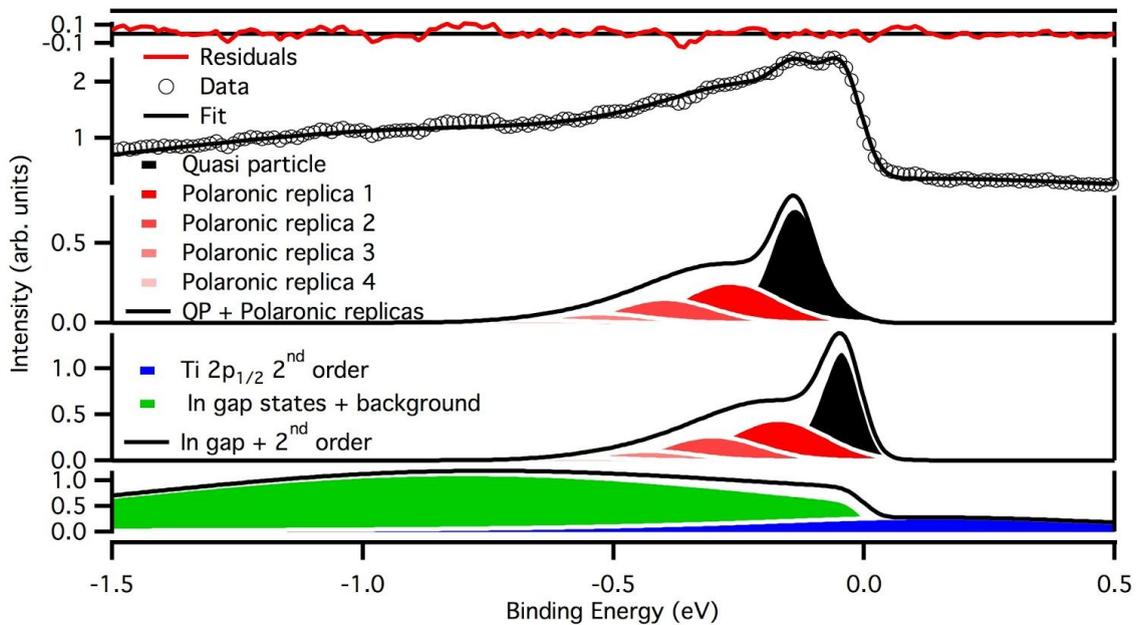

Figure S2: Detailed fit result. From the top: the experimental data and the total result of the fitting procedure, the first QW state composed by its quasiparticle and polaronic tail, the second QW state composed by its quasiparticle and polaronic tail, and the background comprises the in gap states and Ti $2p_{1/2}$ excited by the second order light.

**Thickness dependence - 3 u.c. sample** - Our set of investigated samples comprises also a *n*=3 u.c. sample. Figure S3 shows its electronic structure compared with the calculated one for an equivalent *n*=2.5 sample. The slab used for the calculation is similar to the one used for the *n*=5 sample, and it is composed by 5 u.c. of STO with 4 u.c. of LAO layers on both STO sides, and a vacuum region equivalent to 3 u.c. above LAO.

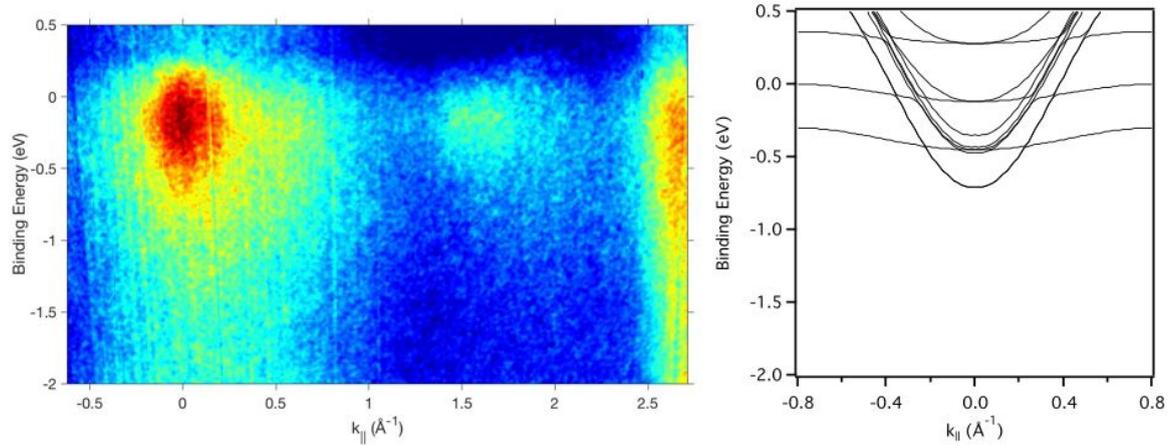

Figure S3: Electronic structure of the *n*=3 u.c. sample (left) and the calculated one for a n=2.5 u.c. sample. In this last the Fermi level is assigned comparing calculations with the experimental band structure.

Experimental features appear too blurry for a quantitative data analysis, however the total bandwidth appears larger compared to thicker samples. This is in accordance with the calculated band structure for thinner samples, where the splitting between the two QW-states increases from 116 meV for the *n*=5 u.c. sample (figure 5 of the main text) to 300 meV for the *n*=2.5 u.c. sample.

Band structure calculated for various STO thickness using the B1-WC functional is shown in figure S4: also in this case, a clear increase of the subband separation decreasing the STO thickness is well captured, even if at a qualitative level.

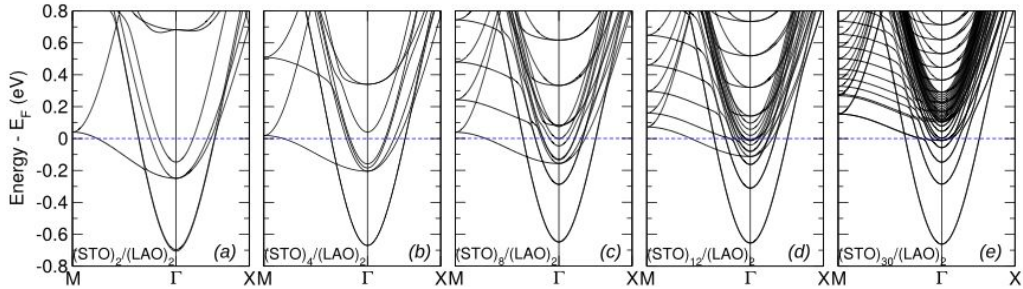

Figure S4: Electronic structure of STO(n)/LAO(2) samples calculated using the B1-WC functional (see following for details).

**Sample growth:** LAO/STO/LAO multilayers have been realized by pulsed laser deposition on commercial STO substrates TiO2 terminated and (001) oriented (from Crystec GmbH). Both LAO and STO were deposited using a laser fluence of ~0.7 J/cm2 and a repetition rate of 1 hz. The first LAO layer was deposited in an atmosphere of 8·10-5 torr of O2 at 800 °C. Later, the temperature was raised at 1100 °C and the pressure decreased to 10-6 torr for the STO deposition. These growth conditions have been optimized to obtain high quality STO films hosting a conducting 2DES. The top LAO layer, 4 u.c., was deposited in 8·10-5 torr of O2 at 800 °C. After the growth the samples were annealed in-situ at 530 °C and 150 torr of O2 for 1 hour and cooled down to room temperature in the same atmosphere. The thickness of the three layers was monitored by reflection high energy electron diffraction (RHEED).

The structural properties of the samples were studied with atomic force microscopy and X-rays diffraction. Experimental θ-2θ scans around the (001) and (002) STO reflections were compared with simulation of the multilayers.

***Ab-initio*** **Calculations: Methodology and technicalities.** - We performed ab-initio calculations for a series of different structures to determine structural and electronic properties of the STO slab under consideration. Two different series of calculations were carried out, corresponding to two different energy functionals: the PSIC and the B1-WC hybrid functional. The former was used to describe the LAO/STO superlattice in neutral (undoped) conditions, the latter to simulate the same structure with electron doping included by construction, and to evaluate the electron charge distribution by Mulliken analysis at varying STO thickness. In the following, we describe the details of the two series of calculations.

**PSIC:** Calculations were carried out with the freeware PWSIC code, implemented in plane waves basis set and ultrasoft pseudopotentials, with 30 Ry cut-off energy, and well-converged **k**-point meshes for reciprocal space integration. All the internal atomic positions are relaxed to the energy minimum with a 0.1 mRy/bohr atomic force threshold. The method is known to accurately reproduce the electronic structure of insulating oxides; as an example, for bulk STO and LAO, PSIC gives 3.0 eV and 5.4 eV band gaps, respectively, which underestimate by less than 10% the measured values [21,22]. For the superlattice the in-plane lattice parameter of STO was assumed, while a full structural relaxation was carried out in the orthogonal direction. To reproduce as closely as possible the features of the actual structure, for the calculation shown in Figure 4 (in the main text) we used a slab composed by 9 and a half u.c. of STO, with 4 u.c. of LAO layers on both STO sides, and a vacuum region equivalent to 3 u.c. above LAO. The additional half layer allows the supercell to have a center of symmetry in the middle of the STO slab. To avoid spurious fields due to periodic boundary conditions, the supercell has a center of symmetry in the middle of the STO slab, so that inversion symmetry along the [001] direction is enforced, giving two symmetry-equivalent STO/LAO interfaces at the two sides of STO, and avoiding spurious fields due to periodic boundary conditions.

**B1-WC:** The CRYSTAL code[1] has been used for the hybrid-functional DFT[2] calculations. They were performed on off-stoechiometric $(STO)_n/(LAO)_2$ ($n$ = 2, 4, 8, 12, 30) superlattices with two *n*-type interfaces (an additional $TiO_2$ plane in STO, and an additional LaO plane in LAO), stacked along the [001] direction and with periodic boundary conditions. Such geometry is adapted to study the fully compensated STO/LAO *n*-type interface, as 0.5 electrons are transferred to each side of the STO as displayed in Figure S5. Moreover, this slab configuration allows to probe the effect of structural confinement on the electronic structure of the 2DES, as the 0.5 electrons remain confined within the half-STO sublattice (ref [30,40] main text). The exchange-correlation energy is modeled with the B1-WC hybrid functional (ref [29] main text), implementing 16% of Hartree-Fock exchange. The hybrid functional approach has also been shown to be qualitatively consistent with the PSIC method (ref [40] main text). A Monkhorst-Pack[3] mesh of 6 × 6 × 1 special k-points of ensures a proper convergence of the total energy at the self-consistent-field level, with a threshold criterion of $10^{-8}$ Ha. The electronic properties are then refined, using a mesh of 12 × 12 × 2 special k-points. A gaussian smearing of the Fermi surface has been set to 0.001 Ha. The basis set used for the different atoms are detailed in Ref. 4 for Ti, Ref. 5 for O, Ref. 6 for Al and Ref. 7 for La. The optimization of the atomic positions are performed with convergence criteria of 1.5 × $10^{-4}$ Ha/Bohr in the root-mean square values of the energy gradients, and 1.2 × $10^{-3}$ Bohr in the root-mean square values of the atomic displacements. The evaluation of the Coulomb and exchange series is determined by five parameters, fixed to their defaults values[1]: 7, 7, 7, 7 and 14.

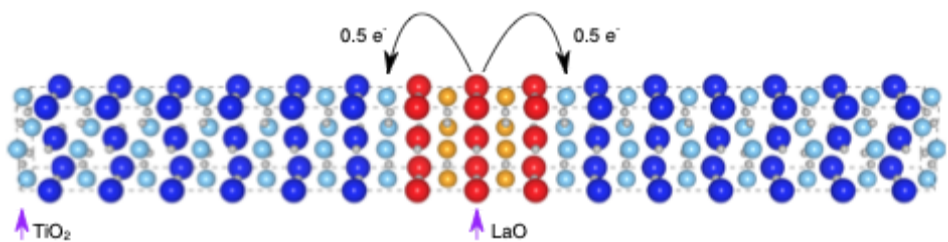

Figure S5: Structure of the (STO)12/(LAO)2 off-stoechiometric superlattice, with the additional TiO2 and LaO planes. The off-stoichiometry of the LAO interlayer ensures that 0.5 electron is transferred to each adjacent TiO2 plane, mimicking a fully-compensated STO/LAO n-type interface.